Channelling study of $La_{1-x}Sr_xCoO_3$ films on different substrates

E. Szilágyi[1*], E. Kótai[1], D. Rata[2], Z. Németh[1], G. Vankó[1]

[1]Institute for Particle and Nuclear Physics, Wigner Research Centre for Physics, P.O.Box 49, H-1525 Budapest, Hungary

[2]Max-Planck-Institut für Chemische Physik fester Stoffe, Nöthnitzer Str. 40, D-01187 Dresden, Germany

**Abstract**

The cobalt oxide system $LaCoO_3$ and its Sr-doped child compounds have been intensively studied for decades due to their intriguing magnetic and electronic properties. Preparing thin $La_{1-x}Sr_xCoO_3$ (LSCO) films on different substrates allows for studies with a new type of perturbation, as the films are subject to substrate-dependent epitaxial strain. By choosing a proper substrate for a thin film grow, not only compressing but also tensile strain can be applied. The consequences for the fundamental physical properties are dramatic: while compressed films are metallic, as the bulk material, films under tensile strain become insulating. The goal of this work is to determine the strain tensor in LSCO films prepared on $LaAlO_3$ and $SrTiO_3$ substrates by pulsed laser deposition using RBS/channelling methods. Apart from the composition and defect structure of the samples, the depth dependence of the strain tensor, the cell parameters, and the volume of the unit cell are also determined. Asymmetric behaviour of the strained cell parameters is found on both substrates. This asymmetry is rather weak in the case of LSCO film grown on $LaAlO_3$, while stronger on $SrTiO_3$ substrate. The strain is more effective at the interface, some relaxation can be observed near to the surface.

---

[*] corresponding author:
e-mail: szilagyi.edit@wigner.mta.hu
Tel: +36 1 392 2222 /ext. 3962

# 1. Introduction

LaCoO$_3$ perovskite and its Sr-doped derivatives have been subject of intensive studies for decades due to their intriguing magnetic and electronic properties that lead to a complex phase diagram including spin-state transition, colossal magnetoresistance and a cooperative ferromagnet – spin-glass and metal-insulator transition [1, 2, 3, 4, 5, 6, 7, 8, 9]. Despite all efforts, the complete understanding of the underlying physics has not been achieved yet.

One of the most important parameters to alter bulk physical properties is the effective Co-O-Co bonding angle, which determines the electronic and magnetic interactions between the cobalt and oxygen ions. This angle depends on the lattice parameters, which are usually altered by hydrostatic or chemical pressure. Preparing thin La$_{1-x}$Sr$_x$CoO$_3$ films on different substrates allows for studies with a new type of perturbation, as the films are subject to well controlled substrate-dependent epitaxial strain. By choosing a proper substrate for a thin film grow, not only compressing but also tensile strain can be applied. The consequences for the fundamental physical properties are dramatic: while compressed LSCO films are metallic, films under tensile strain, with otherwise unchanged parameters such as chemical composition become insulating [10]. Moreover, strain can induce ferromagnetism in the otherwise paramagnetic bulk LaCoO$_3$ material [11, 12]. A few percent of difference in strain leads to many orders of magnitude difference in resistivity and magnetic interactions. This type of tunability of the electronic and magnetic properties by external parameters makes those thin film materials rather interesting for the emerging field of oxide-based electronics.

In the last decades, strain engineering has become an important and necessary technique to improve advanced nanoelectronic devices [13]. Apart from various x-ray [14], neutron [15] and electron [16] diffraction techniques, a powerful method for the non-destructive characterization of strain states in thin layer systems is Rutherford backscattering spectrometry combined with channelling effect (RBS/C) [17, 18].

In this work channelling study have been performed on $La_{1-x}Sr_xCoO_3$ films prepared on $LaAlO_3$ and $SrTiO_3$ substrates to determine either the various types of defects and strain in the films.

## 2. Strain determination from channelling experiments

Strains in crystals are associated with changes in the angles between different crystal directions – except for direction conserving purely hydrostatic strain, i.e., dilatation or compression. From channelling measurements, only the "deviatory" part, $\tilde{\varepsilon}$, defined as the difference between the full strain tensor $\varepsilon$ and its hydrostatic part, can be determined, i.e.,

$$\tilde{\varepsilon} = \varepsilon - trace(\varepsilon/3) \cdot \mathbf{1} = \begin{pmatrix} (2\varepsilon_{11} - \varepsilon_{22} - \varepsilon_{33})/3 & \varepsilon_{12} & \varepsilon_{13} \\ \varepsilon_{12} & (2\varepsilon_{22} - \varepsilon_{11} - \varepsilon_{33})/3 & \varepsilon_{23} \\ \varepsilon_{13} & \varepsilon_{23} & (2\varepsilon_{33} - \varepsilon_{11} - \varepsilon_{22})/3 \end{pmatrix} \quad (1)$$

where **1** is the diagonal unit tensor. The trace of $\tilde{\varepsilon}$ vanishes by construction, i.e., the diagonal elements of $\tilde{\varepsilon}$ are not linearly independent of each other. In cubic system, the strain tensor components, $\varepsilon_{ij}$ can be expressed with the changes in the angle between the direction of [001] and [h,k,l], $\delta_{hkl}$ (in radian) in the following way [18]:

$$\varepsilon_{11} - \varepsilon_{33} = \delta_{101} + \delta_{\bar{1}01};$$

$$\varepsilon_{13} = (\delta_{\bar{1}01} - \delta_{101})/2;$$

$$\varepsilon_{22} - \varepsilon_{33} = \delta_{011} + \delta_{0\bar{1}1} ; \tag{2}$$

$$\varepsilon_{23} = (\delta_{0\bar{1}1} - \delta_{011})/2 \text{ and}$$

$$\varepsilon_{12} = \frac{3}{\sqrt{8}}(\delta_{111} - \delta_{\bar{1}11})$$

Let us note that H. Trinkaus and his co-workers presented an excellent paper [18] where they gave all the required relations between the strain induced angle changes and the components of the strain tensor for general crystalline layer systems of reduced symmetry compared to the basic (cubic) crystal.

## 3. Experimental

$La_{0.7}Sr_{0.3}CoO_3$ films were grown on single crystalline substrates of $SrTiO_3$ (001) (lattice parameter a=3.901 Å) and $LaAlO_3$ (001) (a=3.789 Å) substrates with a size of $5 \times 5$ mm$^2$, by pulsed laser deposition (KrF 248 nm) from a stoichiometric target. The deposition temperature and the oxygen background pressure were 650 °C and $3.5 \times 10^{-1}$ mbar, respectively. The films were cooled down in 600 mbar of oxygen.

In order to perform proper orientation of the crystals using RBS/C a small part of the film (~ 1 mm from the edge of the sample) was removed by ion sputtering using 1 keV Ar using high incident angle of 80°. The glancing angle of incidence is necessary to avoid the defect accumulation, while for reducing the atomic mixing low ion energy is to be used [19].

Composition, defect structure and strain tensor were determined by RBS combined with channelling technique. RBS/C analysis was performed using the 5 MV Van de Graaff accelerator at the Institute for Particle and Nuclear Physics, Wigner Research Centre for Physics, Hungarian Academy of Sciences.

The samples were fastened to a sample holder of a scattering chamber containing a two-axes goniometer. During the experiments the vacuum in the chamber was better than $1 \times 10^{-4}$ Pa using liquid $N_2$ traps along the beam path and around the sample. The ion beam of $^4He^+$ was collimated with 2 sets of four-sector slits to the necessary dimensions of 0.5 x 0.5 mm$^2$. An ion current of typically 10 nA was measured by a transmission Faraday cup [20]. The measured spectra were collected with a dose of 4 µC. The RBS measurements were performed with an ORTEC detector with a solid angle of 4.15 msr. The energy calibration of the multichannel analyzer was determined using known peaks and edges of Au, Si and C.

To determine the sample composition 2000 keV He RBS was performed at tilt angles of 7º, 45º and 56º, respectively. In order to avoid the channelling, the samples were rotated during the measurements. To study the defect structure of the samples RBS/channelling experiments in channel <001> were used with beam energies of 1500, 2000, 2500 keV. The elemental composition and the defect structures of the samples were evaluated from the spectra taken on a sample with the same layer structure using the RBX program [21, 22]. The simulation of crystalline structures, including channelling and the effect of extended and point defects, is usually not handled in analytic codes [23]. At least from the most used simulation codes that took part in IAEA comparison [24, 25] RBX the only one, which is able to include models for given structures and types of defects that can be used for data analysis.

The strain tensors were determined from angular scans around $<111>$, $<\bar{1}11>$ directions through $(1\bar{1}0)$, $(\bar{1}\bar{1}0)$ planes, $<011>$, $<0\bar{1}1>$ directions through $(100)$ plane, and $<101>$, $<\bar{1}01>$ directions through $(010)$ plane. The angle difference between the channelling direction of the layer and the substrate, $\delta$, can be determined with a high accuracy of 0.01°. The components of the strain tensor for cubic crystal can be deduced using the relations between changes in angles of channelling crystal directions [18].

## 4. Results and discussion

Fig 1 shows typical random and channelled RBS spectra taken on the LSCO samples grown on LaAlO$_3$ and SrTiO$_3$ substrates. The evaluation of the spectra of each LSCO film results in an elemental composition of La$_{0.7}$Sr$_{0.3}$Co$_{0.9}$O$_3$ with a layer thickness of 40 nm. A non-single-crystalline layer with a thickness of 2.5 nm was found at the surface of the sample. In the epitaxial layers point defect (9% and 1%, respectively) and strain have to be assumed to simulate all the channelled and random spectra taken with beam energies of 1500, 2000 and 2500 keV using RBX code [21, 22].

La$_{0.7}$Sr$_{0.3}$Co$_{0.9}$O$_3$ films grown on different substrates are subjected to compressive (LaAlO$_3$) and tensile (SrTiO$_3$) strain due to lattice mismatch. To determine the angular scans i.e., normalized yield as a function of the incident angles, two range of interest are chosen, as shown in Fig 1a and 1c. The areas at the surface position of La correspond to the La yield in the films, while the black regions (around channel number 400) correspond to the yield of La and Sr for the case of LaAlO$_3$ and SrTiO$_3$, respectively. From the angular scans the angle difference between the layer and substrate can be determined as shown in Fig. 2. From the detailed channelling data, $\delta_{hkl}$ can be revealed and therefore the strain tensor can be determined via eq. 2.

Dividing the La range of the interest of the film into three parts, the angular scans, $\delta_{hkl}$ show depth dependent behaviour. In Figure 3 the angular scan data of $<\bar{1}11>$ direction is shown in LSCO film on LaAlO$_3$ substrate. From the observed differences in the positions of the dip of the scans, the strain tensor can be determined as a function of the depth.

The cell parameters (x, y, z) of the strained LSCO films as a function of depth can be also calculated from the strain tensor as shown in Figure 4. In case of the LaAlO$_3$ substrate the cell is compressed in a slightly asymmetric way with respect to the x and y directions (in plane directions), and this compression results a dilatation in z direction (out of plane direction). A stronger asymmetric behaviour of the dilatation can be observed in the case of the film grown on SrTiO$_3$ substrate, where the cell is dilated in y direction more effectively than in x direction; this dilatation results again in a compression in z direction. The strain is more effective at the interface, some relaxation can be observed at the surface. Assuming that the deformed LSCO cells follow the lattice of the substrate, the volume of the unit cell of LSCO can be easily calculated and it is found to be 55.96 ± 0.09 and 57.87 ± 0.09 Å$^3$ grown on LaAlO$_3$ and SrTiO$_3$ substrates, respectively. The uncertainty of the volume of the unit cell is calculated taking into account the accuracy of the angle resolution of 0.01°. The unit-cell volume of LSCO formed on various substrates follows similar behaviour as it is calculated from the lattice parameters determining via x-ray diffraction data from ref. [10]. Similar tendency is observed for LaCoO$_3$ films grown on various substrates [11].

## 5. Conclusions

La$_{1-x}$Sr$_x$CoO$_3$ films prepared on LaAlO$_3$ and SrTiO$_3$ substrates were investigated by RBS/channelling methods. Apart from the composition and defect structure of the samples, from the detailed channelling data, the depth dependence of the strain tensor, the cell parameters, and the volume of the unit cell can be determined.

Asymmetric behaviour of the cell parameters was found near to the substrate. This asymmetry is rather weak in the case of LSCO film on LaAlO$_3$ substrate, where the cell is compressed in x and y directions, resulting a dilatation in z direction. A more definite asymmetry is found in the case of SrTiO$_3$ substrate, where the LSCO cell is dilated in y direction more effective than in x direction; while it was compressed in z direction. The strain is more effective at the interface, some relaxation can be observed near to the surface. Assuming that the deformed LSCO cells follow the lattice of the substrate, the unit-cell volume of LSCO is found to be 55.96 ± 0.09 and 57.87 ± 0.09 Å$^3$ grown on LaAlO$_3$ and SrTiO$_3$ substrates, respectively.

**Acknowledgments**

The support of ERC-StG-259709 and the Hungarian grant OTKA K 72597 is acknowledged. The authors thank A. Sulyok for removing the part of the films by ion sputtering. RBS/C experiments were carried out in the frames of the Hungarian Ion-beam Physics Platform.

**Figures for the electronic version**

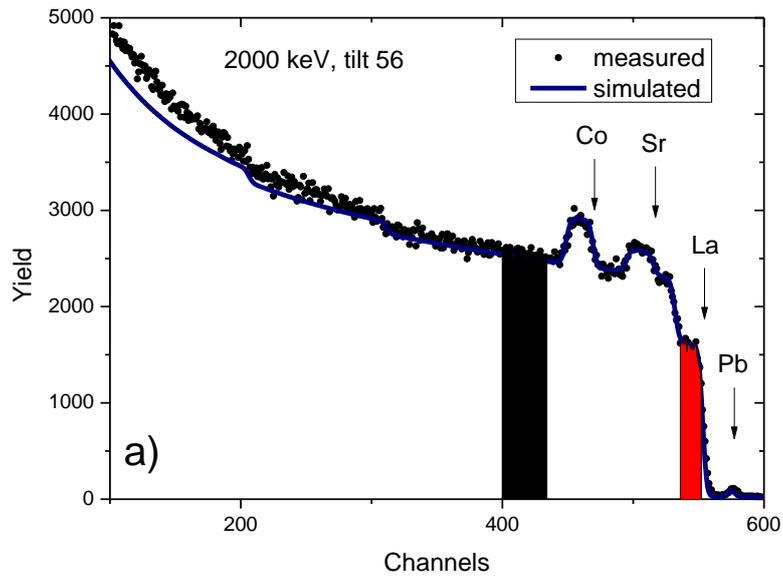

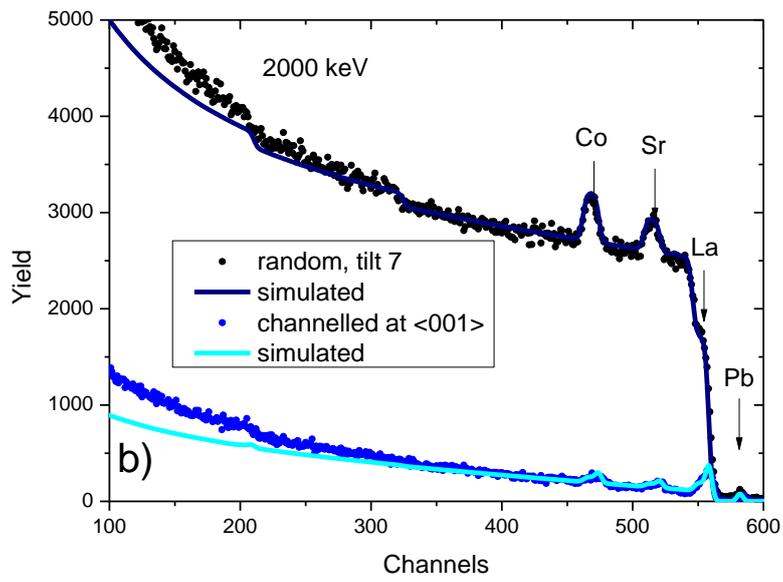

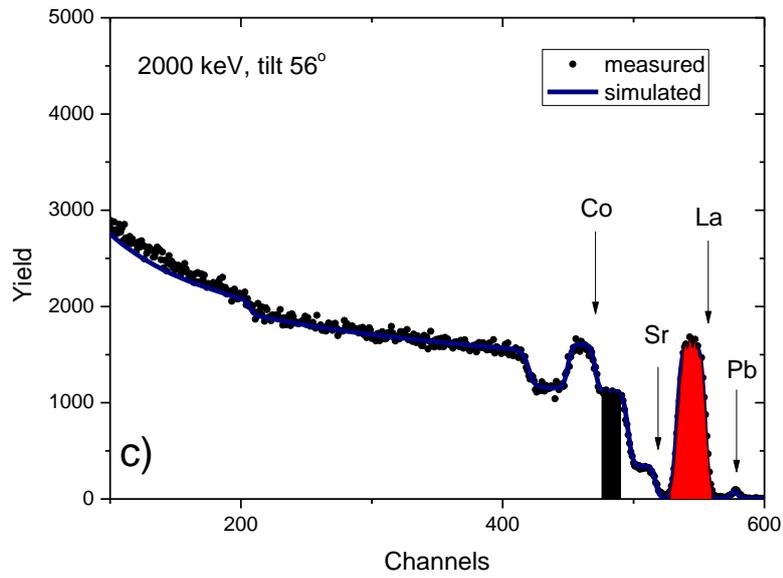

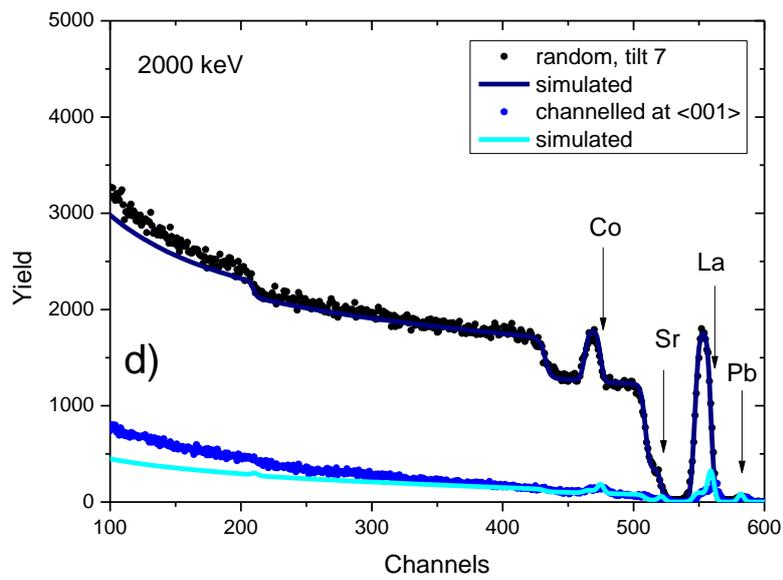

Figure 1. Measured and simulated RBS (a,c) and RBS/channelling (b,d) spectra using beam energy of 2000 keV taken on $La_{1-x}Sr_xCoO_3$ $x = 0.3$ film grown on $LaAlO_3$ (a, b) and $SrTiO_3$ (c, d) substrates.

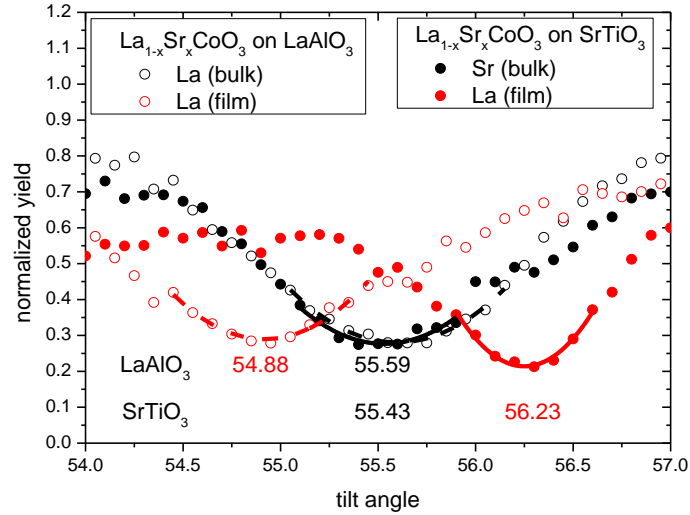

Figure 2. Angular scans around $<\bar{1}11>$ direction normalized to the random level. Fig 1 a) and c) show the integrals that correspond to the substrates (black areas) and to the La (areas indicated with a La sign) in the LSCO films. A significant angle difference can be observed between LSCO films grown on $LaAlO_3$ and $SrTiO_3$ substrates due to the different lattice mismatch induced stresses (compressive on $LaAlO_3$ and tensile strain on $SrTiO_3$ substrates), which leads to different strained layers. The fitted positions of the minimum of the parabolas are also given.

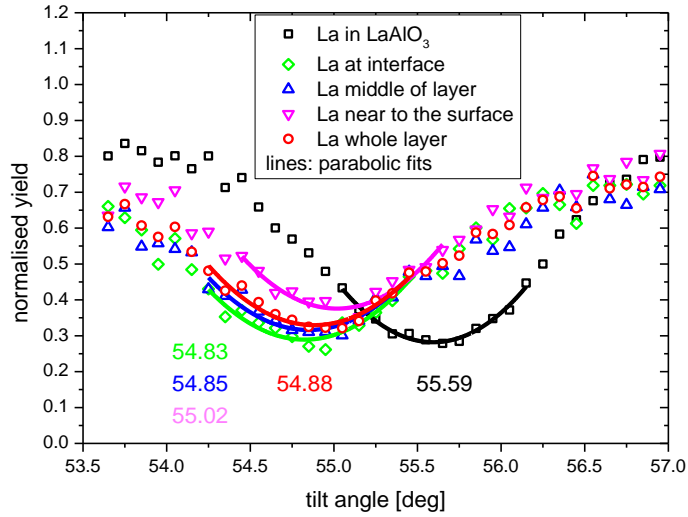

Figure 3. Determination of the depth dependence in the angular scan data of $<\bar{1}11>$ direction, shown in the case of the LSCO film on LaAlO$_3$ substrate. La integral of the film (see the La areas in fig.1 a and c) is divided to three parts (◇ La at the interface, △ in the middle of the layer and ▽ near to the surface). For comparison, the La in the whole layer (○) and the substrate (□) are also plotted. The parabolic fits to the data (lines) and the positions of the minima are also shown.

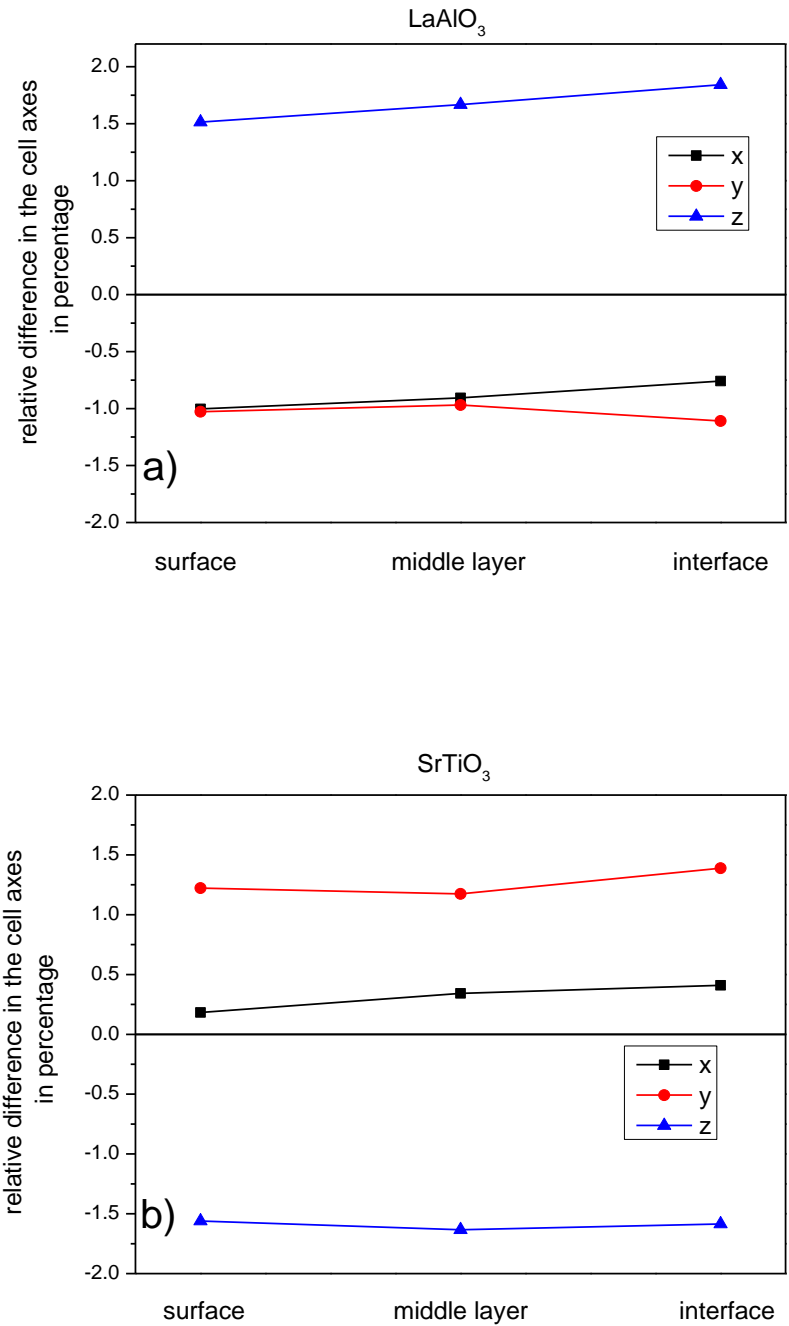

Figure 4. Relative differences in the cell axes of LSCO films grown on a) LaAlO$_3$ and b) SrTiO$_3$ substrates are plotted in percentage as a function of depth. The uncertainty of the data points (± 0.04) is smaller than the size of the symbols.